\documentclass[a4paper]{jpconf}
\usepackage{graphicx}
\begin{document}
\title{Mass-radius constraints for the neutron star EoS - Bayesian analysis}

\author{A.~Ayriyan$^{1}$, D.E.~Alvarez-Castillo$^{2,3}$, D.~Blaschke$^{3,4,5}$, H.~Grigorian$^{1,6}$}
\address{$^1$Laboratory for Information Technologies, JINR Dubna, 141980 Dubna, Russia}
\address{$^2$Bogoliubov Laboratory for Theoretical Physics, JINR Dubna, 141980 Dubna, Russia}
\address{$^3$Universidad Aut\'{o}noma de San Luis Potos\'{i}, S.L.P. 78290 M\'{e}xico, Mexico}
\address{$^4$Institute of Theoretical Physics, University of Wroc{\l}aw, 50-204 Wroclaw, Poland}
\address{$^5$National Research Nuclear University (MEPhI), 115409 Moscow, Russia}
\address{$^6$Department of Physics, Yerevan State University, Yerevan, Armenia}
\ead{ayriyan@jinr.ru}

\begin{abstract}
We suggest a new Bayesian analysis (BA) using disjunct M-R constraints for 
extracting probability measures for cold, dense matter equations of state (EoS). 
One of the key issues of such an analysis is the question of a
deconfinement transition in compact stars and whether it proceeds as a 
crossover or rather as a first order transition. 
We show by postulating results of not yet existing radius measurements for the known pulsars with
a mass of $2M_\odot$ that a radius gap of about 3 km would clearly select an EoS with a strong first order phase transition as the most probably one.   
This would support the existence of a critical endpoint in the QCD phase diagram under scrutiny in 
present and upcoming heavy-ion collision experiments. 
\end{abstract}

\section{Introduction: EoS with a QCD phase transition and hybrid stars}
There is a unique relationship between the  EoS of cold dense matter and the sequence of compact star 
configurations in the mass-radius (M-R) diagram provided by the Tolman-Oppenheimer-Volkoff equations. 
It can be used to quantify the likelihood of EoS models by Bayesian analyses using a selection of mass and/or radius measurements as priors  \cite{Steiner:2010fz}.
In this work we discuss a family of hybrid neutron stars described by an EoS with a first-order phase transition that is obtained from a Maxwell construction between hadronic and quark matter branches. 
Our approach to the hybrid EoS captures important physical features:
\begin{itemize}
 \item excluded volume corrections in the hadronic EoS, 
 \item multi-quark interactions for quark matter EoS. 
\end{itemize}
The excluded volume corrections at the hadronic level take into account the inner structure of the nucleons which results in Pauli blocking due to quark exchange interactions. 
This can be seen as a precursor of the quark delocalization which is the expected mechanism of the deconfinement transition \cite{Ropke:1986qs} that is still intensified by the simultaneous 
partial chiral restoration \cite{Blaschke:2016}. 
The excluded volume modification is implemented for baryon densities exceeding saturation 
$n_{\textmd{sat}}=0.16$ fm$^{-3}$ 
(the density in the interior of atomic nuclei) by considering the available volume $\Phi_N$ for the motion of nucleons at a given density $n$ as defined in~\cite{Typel:2015xxx}

 \begin{equation}
    \Phi_N=\left\{
                \begin{array}{lll}
                  1~,& \textmd{if} &n \leq n_{\textmd{sat}}\\
                  \exp[-{v\vert v \vert}(n-n_{\textmd{sat}})^{2}/2]~, & \textmd{if} &n > n_{\textmd{sat}}~,
                \end{array}
              \right.
              \label{vex}
  \end{equation}
where $v=16\pi r_N^3/3$ 
is the van-der-Waals excluded volume that is related to the nucleon hard-core radius $r_N$.
Note that in the form (\ref{vex}) assigning positive (negative) values for $v$ leads to a stiffening (softening)
of the original EoS for which we adopt DD2 \cite{Typel:1999yq} with parameters from \cite{Typel:2009sy}. 
Here, we will consider only positive values of excluded volume and introduce for them the dimensionless parameter $p=10\times v[{\rm fm}^3]$, varying it in the range from $0$ to $80$.

\begin{figure}[!htb]
\begin{tabular}{cc}
 \includegraphics[height=.35\linewidth]{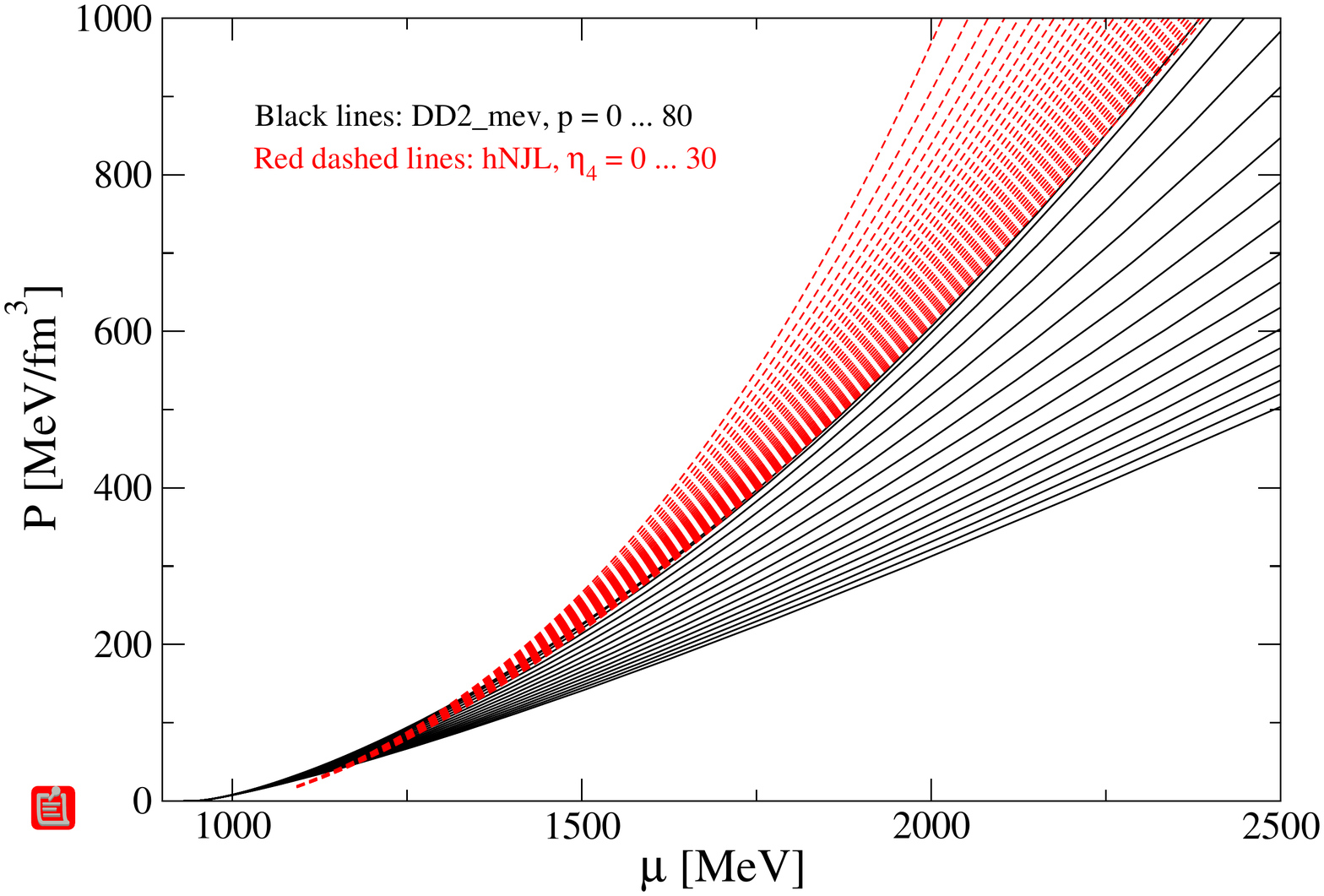}&
 \includegraphics[height=.35\linewidth]{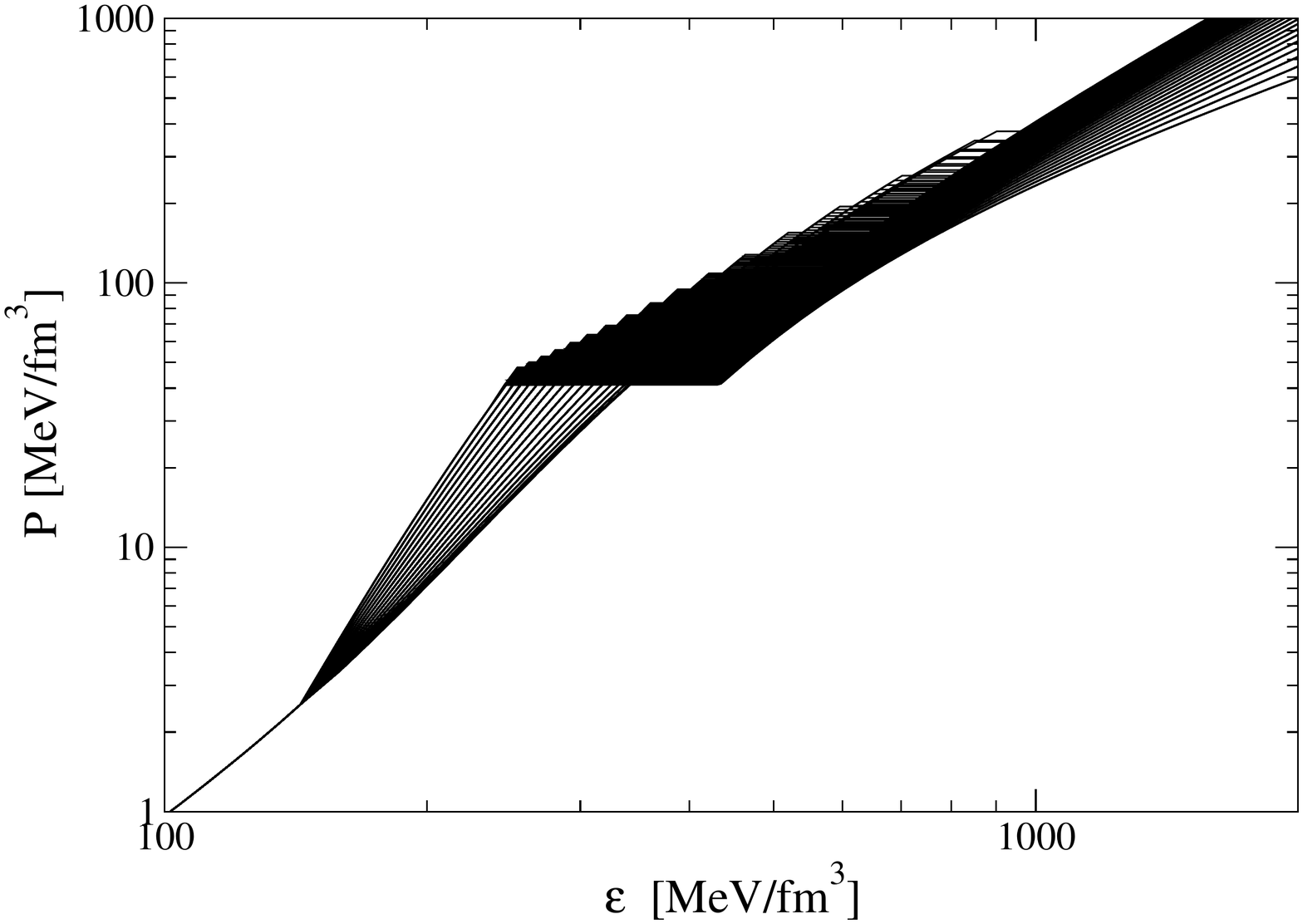}\\
 \includegraphics[height=.35\linewidth]{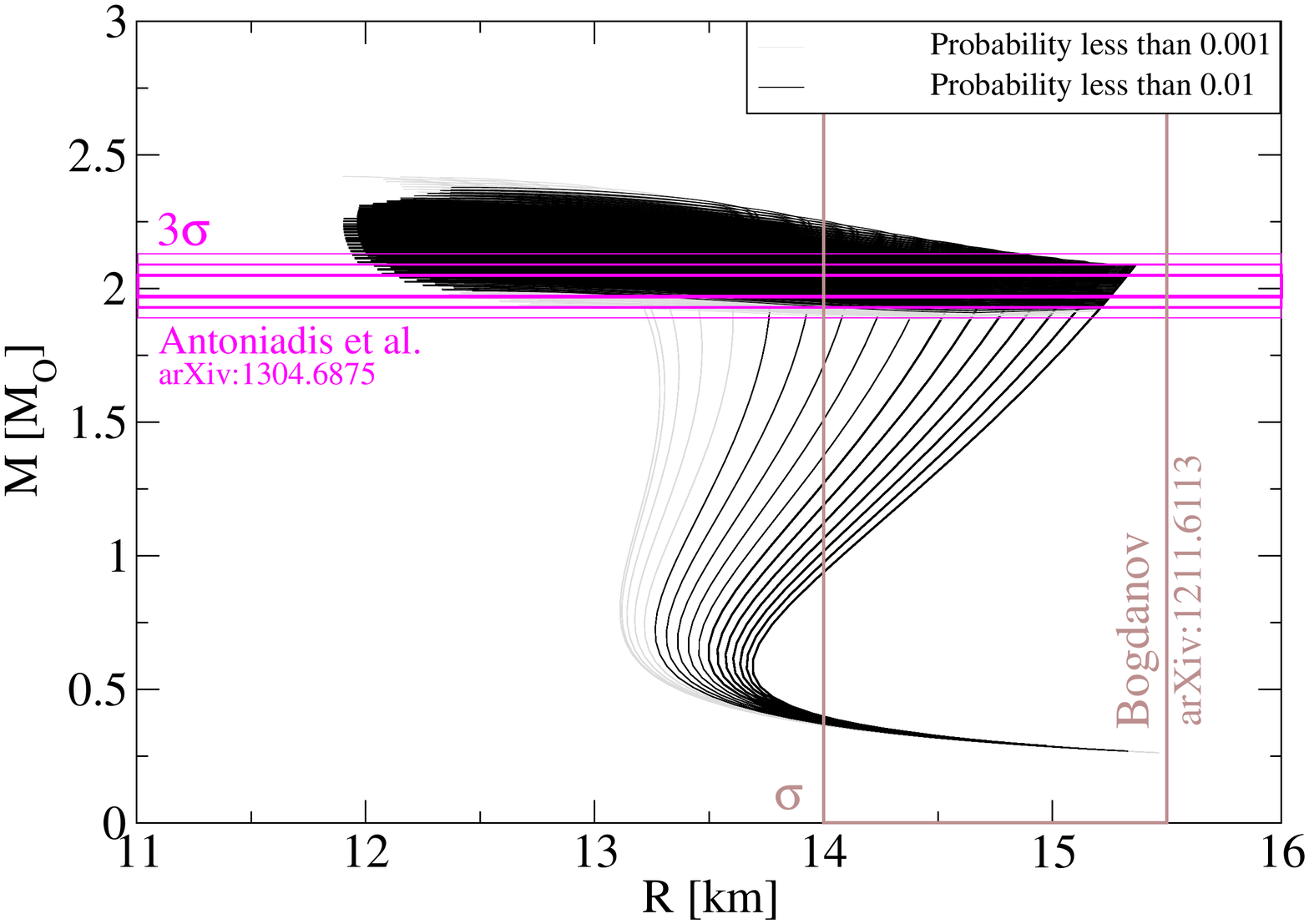}&
 \includegraphics[height=.35\linewidth]{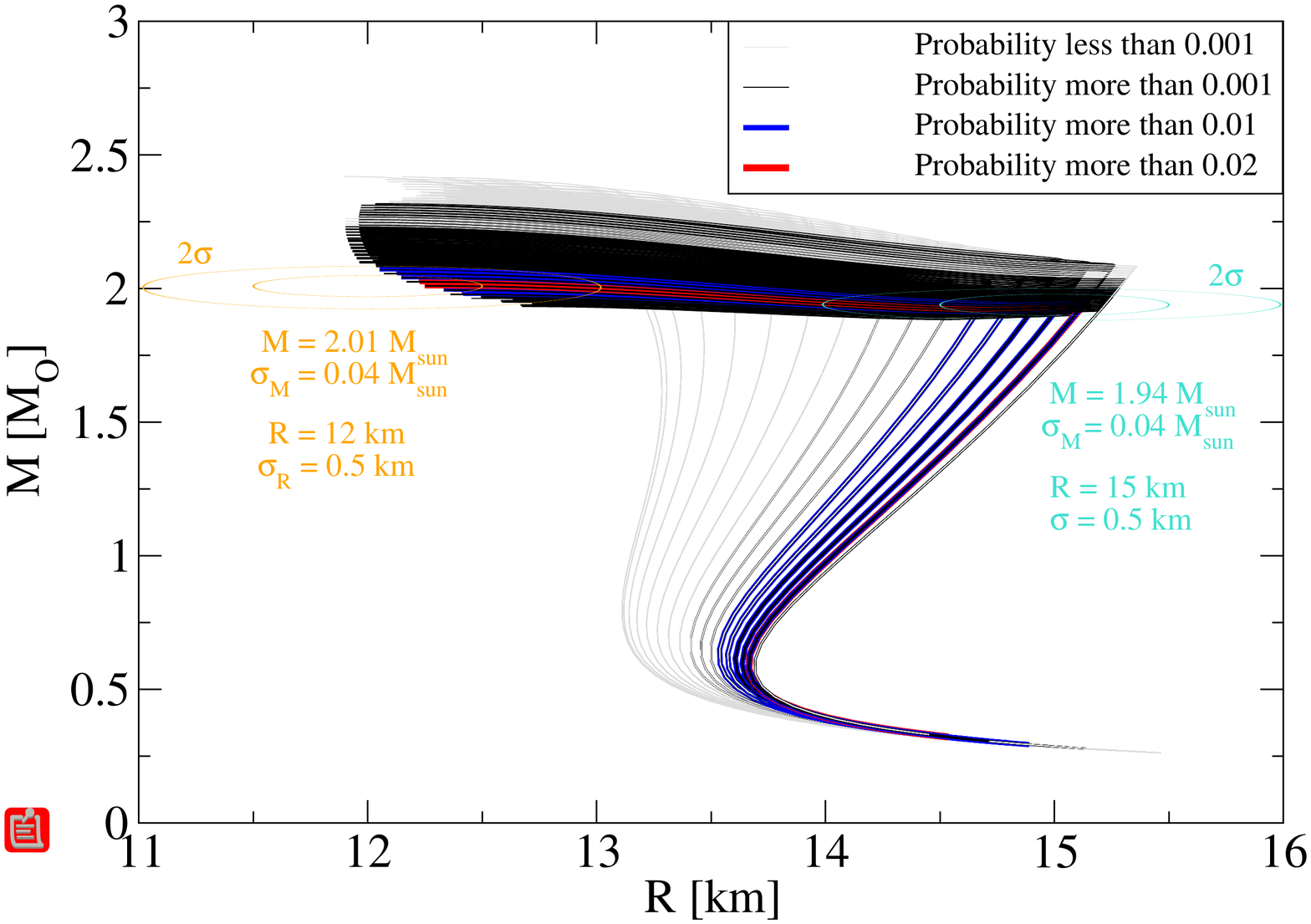}\\
 \end{tabular}
 \caption{Hybrid EoS scheme for sets of two parameters 
 $p$ and $\eta_4$ corresponding to hadronic and quark matter~\cite{Alvarez-Castillo:2015via}. 
 Upper panels: 
 pressure vs. chemical potential (left) for hadronic (DD2\underline{ }mev, black solid lines) and 
 quark matter (hNJL, red dashed lines); 
 pressure vs. energy density with Maxwell constructions of first order phase transitions  (right).
 The two EoS parameters are varied in the range  $(p,\eta_4)=(0\dots 80, 0\dots 30)$.
 Lower panels: The corresponding mass-radius relations are compared to existing mass 
 and radius measurements (left) and to ficticious (right) priors for which it is assumed that 
 for the two objects with measured high masses, PSR J0348+0432~\cite{Antoniadis:2013pzd} and 
 PSR J1614-2230~\cite{Demorest:2010bx} also their radii could be measured with a $1\sigma$ 
 accuracy of $\sim 0.5$ km and turn out to be significantly different at the $3\sigma$ level with mean 
 values of $15$ km and $12$ km, resp.}
 \label{AHP_Scheme}
\end{figure}

We consider the multi-quark interactions as they were introduced by~\cite{Benic:2014iaa} and succesfully applied to massive hybrid star twins in~\cite{Benic:2014jia}. 
It is expected that quark matter will still be strongly interacting at the energy densities present in neutron star interiors.
Those fall in the range where in lattice QCD simulations at finite temperatures the transition from hadronic to quark matter is obtained. 
Therefore the introduction of the 8-quark vector channel interaction may be justified.
In accordance with Refs.~\cite{Benic:2014iaa,Benic:2014jia}, we denote the dimensionless coupling constant of this channel by $\eta_4$ and vary it in the range from $0$ to $30$.
Increasing the value of this parameter has the effect of stiffening the EoS at higher densities, which allows to support hybrid stars as massive as 2~M$_{\odot}$ and thus fulfilling the constraint from recent observations \cite{Demorest:2010bx,Antoniadis:2013pzd}, see Fig.~\ref{AHP_Scheme}.

\section{Observational constraints and Bayesian analysis}
We want to compare the theoretical results for masses and radii of hybrid star sequences with suitable constraints from neutron star observations in order to conclude for the likeliness of the underlying EoS model. 
We consider two cases, a real (A) and a ficticious (B) set of measurements:
\begin{enumerate}  
\item case A
\begin{itemize}
\item a maximum mass constraint from PSR J0348+0432~\cite{Antoniadis:2013pzd}, 
\item a radius constraint from the nearest millisecond pulsar and
PSR J0437-4715~\cite{Bogdanov:2012md}, 
\end{itemize}
\item case B
\begin{itemize}
\item a radius measurement of $R=12 \pm 0.5$ km for PSR J0348+0432 with its known mass,
\item a radius measurement of $R=15 \pm 0.5$ km for PSR J1614-2230 with its known mass,
\end{itemize}
\end{enumerate}
which both are shown in the lower panels of Fig.~\ref{AHP_Scheme}.

For the BA of the most probable EoS for given prior from (real or ficticious) observations, 
we start by defining a vector of free parameters $\overrightarrow{\pi}(p,\eta_4)$, 
which correspond to all the possible models with phase transition from nuclear to quark matter using the EoS described above. The way we sample these parameters is
\begin{equation}
\label{pi_vec}
\pi_i = \overrightarrow{\pi}
\left(p(k),\eta_4(l)\right),
\end{equation}
where $i = 0, 1,\dots, N-1$ with $N = N_1\times N_2$ 
such that $i = N_2\times k + l$ and 
$k = 0, 1, \dots, N_1-1$, $l = 0, 1, \dots, N_2-1$, with $N_1$ and $N_2$ as the total number of parameters $p_k$ and $\eta_{(4)l}$ respectively.
The goal is to find the set $\pi_i$ corresponding to an EoS and thus a 
sequence of configurations which contains the most probable 
one based  on the given constraints using BA.
For initializing the BA we propose that {\it a priori} each vector of 
parameters $\pi_i$ has the same probability: 
$P\left(\pi_i\right) = 1/N$ for $\forall i$.
We can calculate probability of $\pi_i$ using Bayes' theorem 
\cite{Blaschke:2014via}
\begin{equation}
\label{pi_apost}
P\left(\pi_i\left|E\right.\right) = 
\frac{{P\left(E\left|\pi_i\right.\right)P\left(\pi_i\right)}}
{{\sum\limits_{j=0}^{N-1}
P\left(E\left|\pi_j\right.\right)P\left(\pi_j\right)}}.
\end{equation}


\begin{figure}[!th]
\begin{tabular}{ccc}
 \includegraphics[height=.40\linewidth]{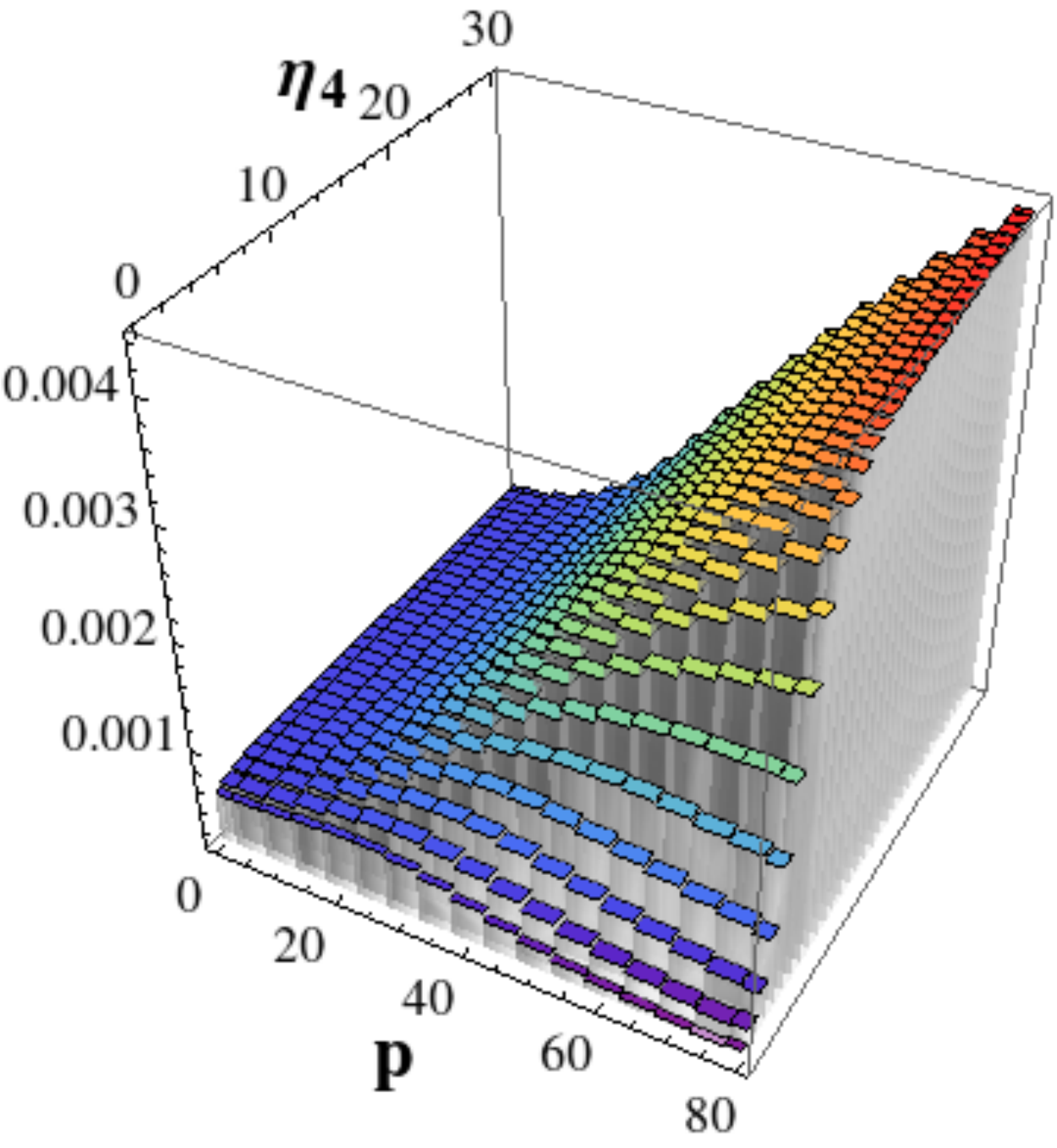}&\qquad&
 \includegraphics[height=.40\linewidth]{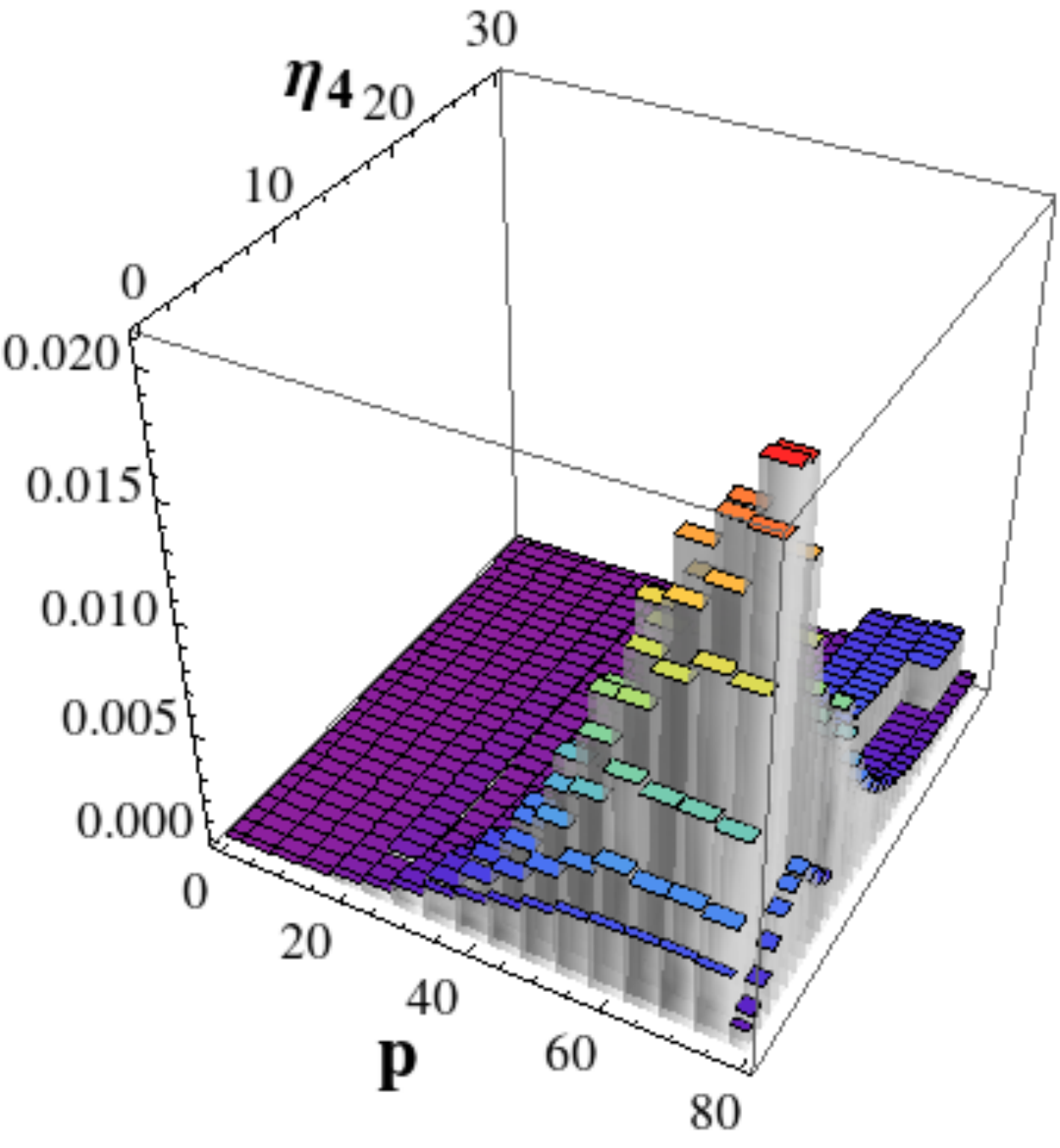}\\
\end{tabular}
 \caption{LEGO plots for Bayesian analysis in the plane of two parameters 
$\left(p, \eta_{4} \right)$; for their definition see text. 
The left panel uses the priors of case A: the mass measurement by Antoniadis et al. \cite{Antoniadis:2013pzd} and the radius measurement by Bogdanov \cite{Bogdanov:2012md} as given in the lower left panel of Fig.~\ref{AHP_Scheme}. 
The result in the right panel considers the priors of case B, with two ficticious radius measurement, as in the lower right panel of Fig.~\ref{AHP_Scheme}.}
 \label{FIG:flowPRS}
\end{figure} 

The results of our BA for the two cases defined above are shown in the two panels of 
Fig.~\ref{FIG:flowPRS}. For the case A, due to the constraint of a large radius, the stiffest possible EoS among the accessible parameters was selected as the most probable hybrid star EoS. 
In particular, largest values of the parameters $p$ and $\eta_4$  that would correspond to hybrid star branches connected with hadronic ones, in the classification of   \cite{Alford:2013aca}.
For case B, the high-mass twin star solution is preferred which corresponds to a disconnected hybrid star branch ("third family") and disselects large values of the $\eta_4$ parameter. 
Herewith we could demonstrate the usefulness of the Bayesian analysis method. It allows to check the discriminating power of a certain experimental result before the actual measurement was made. 
In this sense the discovery potential of future programmes for compact star  observations, like NICER or SKA  could be quantified.

\section{Conclusions}
To test the interesting possibility that a strong first-order phase transition 
might occur in massive neutron stars of $2~M_\odot$ 
\cite{Antoniadis:2013pzd} leading to the appearance of a 
``third family'' of hybrid stars (mass twins) at this mass, one should 
perform radius measurements for these massive neutron stars 
\cite{Blaschke:2014via}.
This possibility offers bright prospects for future observational campaigns and
bears the chance to ``prove'' the existence of a critical point 
\cite{Benic:2014jia,Alvarez-Castillo:2015via,Blaschke:2014via,Blaschke:2013ana,Alvarez-Castillo:2013cxa,Alvarez-Castillo:2014nua,Alvarez-Castillo:2014dva} 
in the QCD phase diagram from astrophysics.

\section{Acknowledgments}
We acknowledge fruitful discussions with M. C. Miller and J. Tr\"umper.
We thank S. Typel and S. Benic for providing us with their EoS data.
D.B., H.G. and D.E.A.-C. profited from support of networking activities by the
COST Action MP1304 ``NewCompStar''.
This work was supported by the 
NCN ``Opus'' programme under contract 
UMO-2014/13/B/ST9/02621.
D.E.A-C., A.A.  and H.G. received support from the 
Bogoliubov--Infeld and the Ter-Antonian--Smorodinsky programmes.  
A.A. acknowledges JINR grant No. 14-602-01.

\end{document}